\twocolumn
\documentclass[conference]{IEEEtran}
\usepackage{cite}
\usepackage[cmex10]{amsmath}
\usepackage{amssymb}
\usepackage{algorithm}
\usepackage{algorithmicx}
\usepackage{algpseudocode}
\usepackage{array}
\usepackage{enumerate}
\usepackage{booktabs}
\usepackage{url}
\usepackage{setspace}
\usepackage{graphicx}
\usepackage{color}

\begin{document}

\title{Improving Crowdsourced Live Streaming with Aggregated Edge Networks}


\author{
  	Chenglei Wu, Zhi Wang, Jiangchuan Liu, Shiqiang Yang\\
  	\{wcl15@mails., wangzhi@sz., yangshq@\}tsinghua.edu.cn, jcliu@cs.sfu.ca
}

\maketitle

\IEEEpeerreviewmaketitle


\begin{abstract}

	Recent years have witnessed a dramatic increase of user-generated video services. In such user-generated video services, \emph{crowdsourced} live streaming (e.g., Periscope, Twitch) has significantly challenged today's edge network infrastructure: today's edge networks (e.g., 4G, Wi-Fi) have limited uplink capacity support, making high-bitrate live streaming over such links fundamentally impossible. In this paper, we propose to let broadcasters (i.e., users who generate the video) upload crowdsourced video streams using \emph{aggregated} network resources from multiple edge networks. There are several challenges in the proposal: First, how to design a framework that aggregates bandwidth from multiple edge networks? Second, how to make this framework transparent to today's crowdsourced live streaming services? Third, how to maximize the streaming quality for the whole system? We design a multi-objective and deployable bandwidth aggregation system \emph{BASS} to address these challenges: (1) We propose an aggregation framework transparent to today's crowdsourced live streaming services, using an edge proxy box and aggregation cloud paradigm; (2) We dynamically allocate geo-distributed cloud aggregation servers to enable MPTCP (i.e., multi-path TCP), according to location and network characteristics of both broadcasters and the original streaming servers; (3) We maximize the overall performance gain for the whole system, by matching streams with the best aggregation paths.
\end{abstract}


\section{Introduction} \label{sec:intro}

Recent years have witnessed a dramatic increase of user-generated video services, enabled by the popularity of online video service and online social network service, and their cross pollination\cite{li2013two}. Among such user-generated video services, \emph{crowdsourced} live streaming(e.g., Periscope, Twitch), which allows individuals to broadcast \emph{live} video streams to millions of users on the move, basks in surge of interest. For instance, users collectively watch over $40$ years worth of live streaming content on Periscope \cite{periscopemilestone}, one of the representative crowdsourced live streaming apps.



\begin{figure}
    \centering
    \includegraphics[width=.7\linewidth]{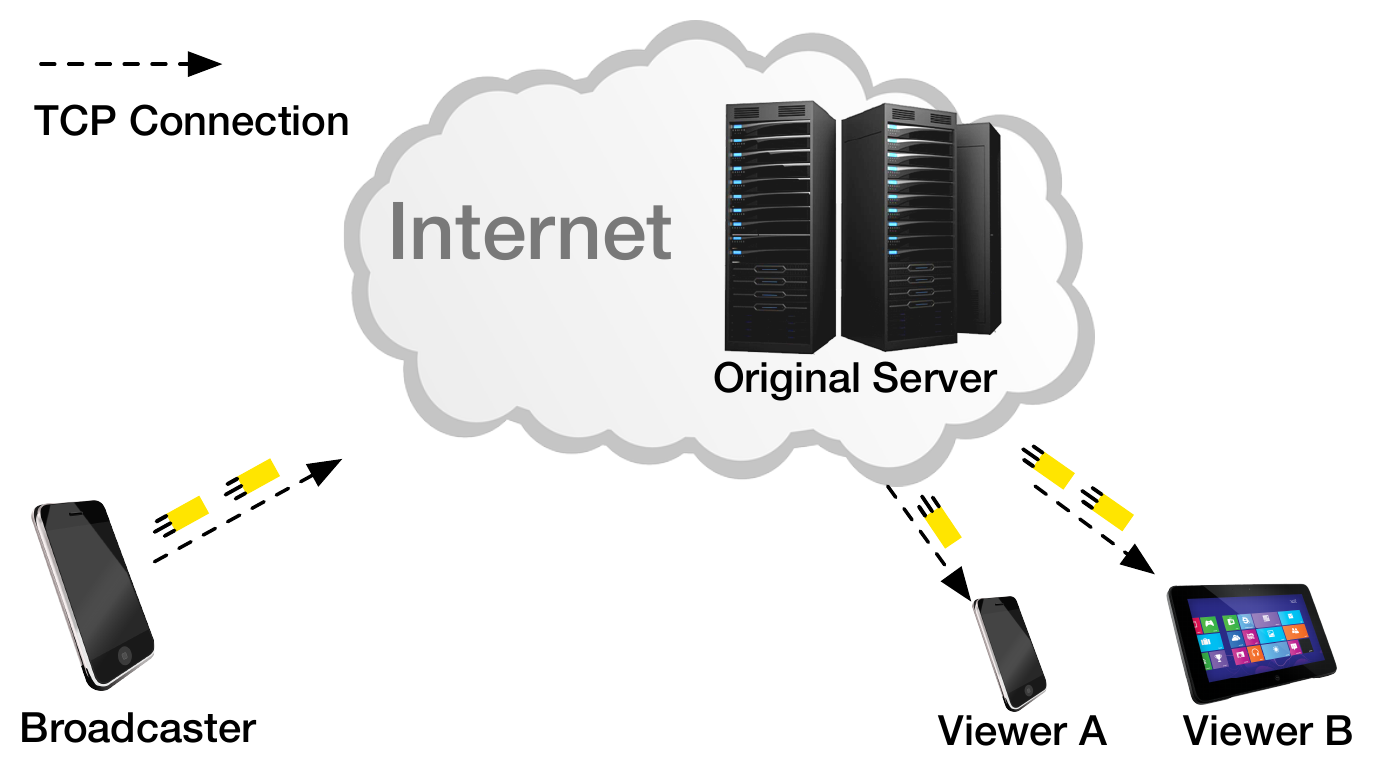}
    \caption{Crowdsourced live video streaming: users upload live video streaming
    via edge networks, e.g., 3G.}
    \label{fig:conventional}
\end{figure}

In live streaming service, we distinguish \emph{broadcasters} (i.e., users who generate and upload the video streaming) and \emph{viewers} (i.e., users who watch the live streaming). Fig. \ref{fig:conventional} illustrates the conventional approach for crowdsourced live streaming: the original stream from a broadcaster will be transferred to streaming servers, usually via TCP; then the video stream will be delivered to the viewers by the streaming servers.
Although many researchers have been working on this topic, most of these studies \cite{pires2014dash,zhang2015crowdsourced, chen2015crowdsourced} only focused on the video delivery part (i.e., how video data is delivered to users), and left the uploading (i.e., how video data is effectively uploaded to the servers) side unexplored.


Given the fact that today's mobile devices can take high-resolution videos,
e.g., iPhone 6s supports 30 fps 4K High Definition video recording\footnote{
According to Apple, a minute of video with 1080p HD at 30 fps is approximately
130MB and a minute of video with 4K higher resolution is approximately 375MB.},
it is challenging for today' edge network infrastructure to fulfill the large
bandwidth requirement: to effectively online stream a 1080P HD video, the
bandwidth required is at least 5 Mbit/s, and 15 Mbit/s for 4K streaming with
efficient video compression methods. For live broadcasting which is unable to
compress video data efficiently, the bandwidth requirement would be inevitably much larger.

Wi-Fi networks along with cellular networks (e.g., 3G, 4G) and other networks compose
the common edge networks. The bandwidths of current edge
networks, however, often fail to satisfy the requirement:
almost 60\% of users have less than 1 Mbits/s upload bandwidth according to our measurement
study on Wi-Fi quality in representative cities.
One of the reasons is that local carrier often configures the upload rate
much lower than the download rate, since most people have more of a need to download data.
And according to Akamai's research, only 21\% of U.S. homes have more than 15 Mbits/s bandwidths\footnote{http://www.fiercecable.com/story/akamai-only-21-us-homes-have-enough-bandwidth-stream-4k/2015-09-23},
which also confirms that uplink capacity is relatively limited due to the same reason.
The upload capacity of 4G cellular network is insufficient for 4K live streaming and the cost is rather
high. Therefore the gap between bandwidth requirement and reality significantly
challenged today's wireless infrastructure.

\begin{figure*}[t]
    \begin{minipage}{.333\textwidth}
        \centering
        \includegraphics[trim={1cm 7cm 2cm 8cm},clip, width=\linewidth]{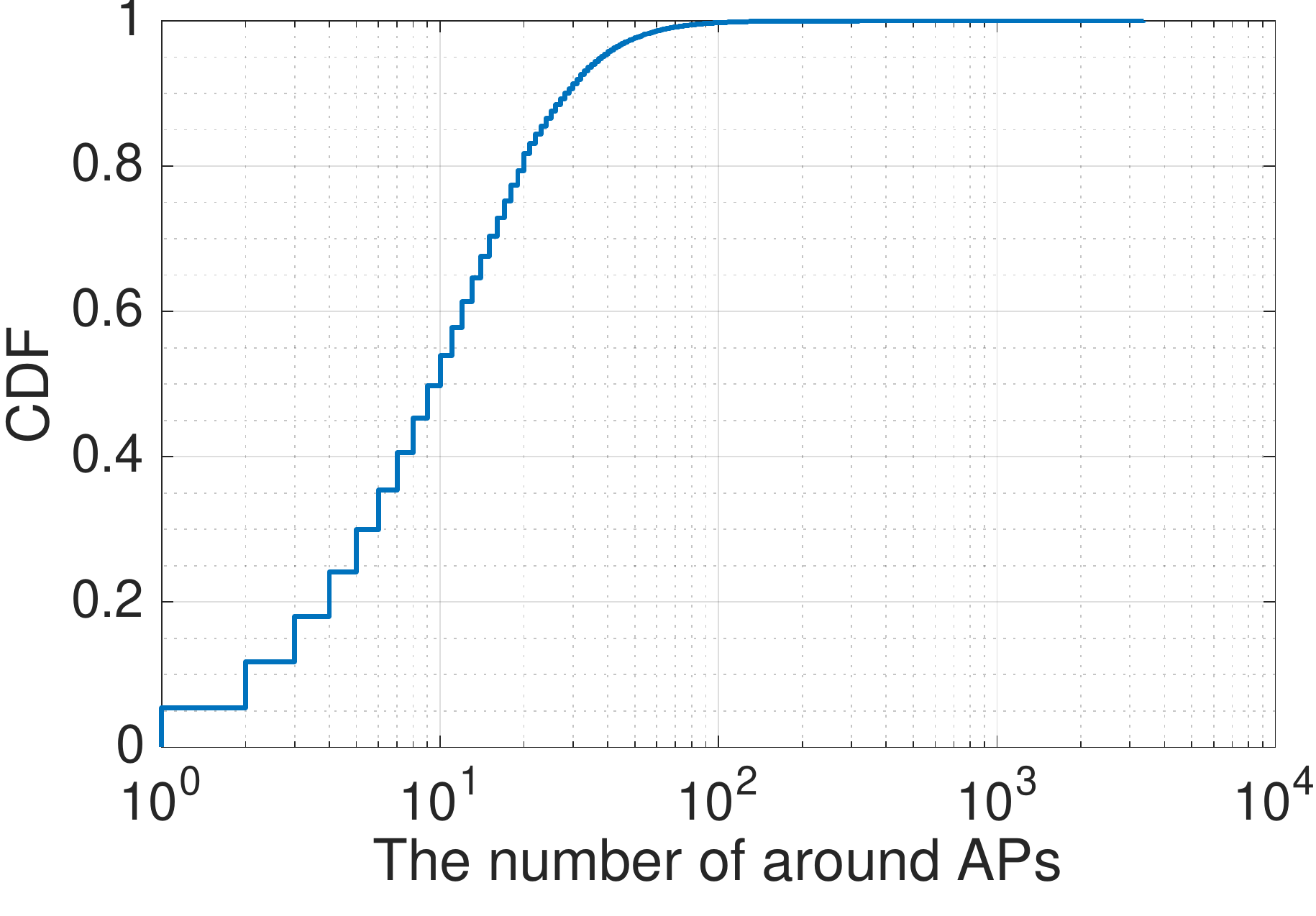}
        \caption{Number of Wi-Fi access point sensed by the devices.}
        \label{fig:aroundAp}
    \end{minipage}
    \begin{minipage}{.333\textwidth}
        \centering
        \includegraphics[trim={1cm 7cm 2cm 8cm},clip, width=\linewidth]{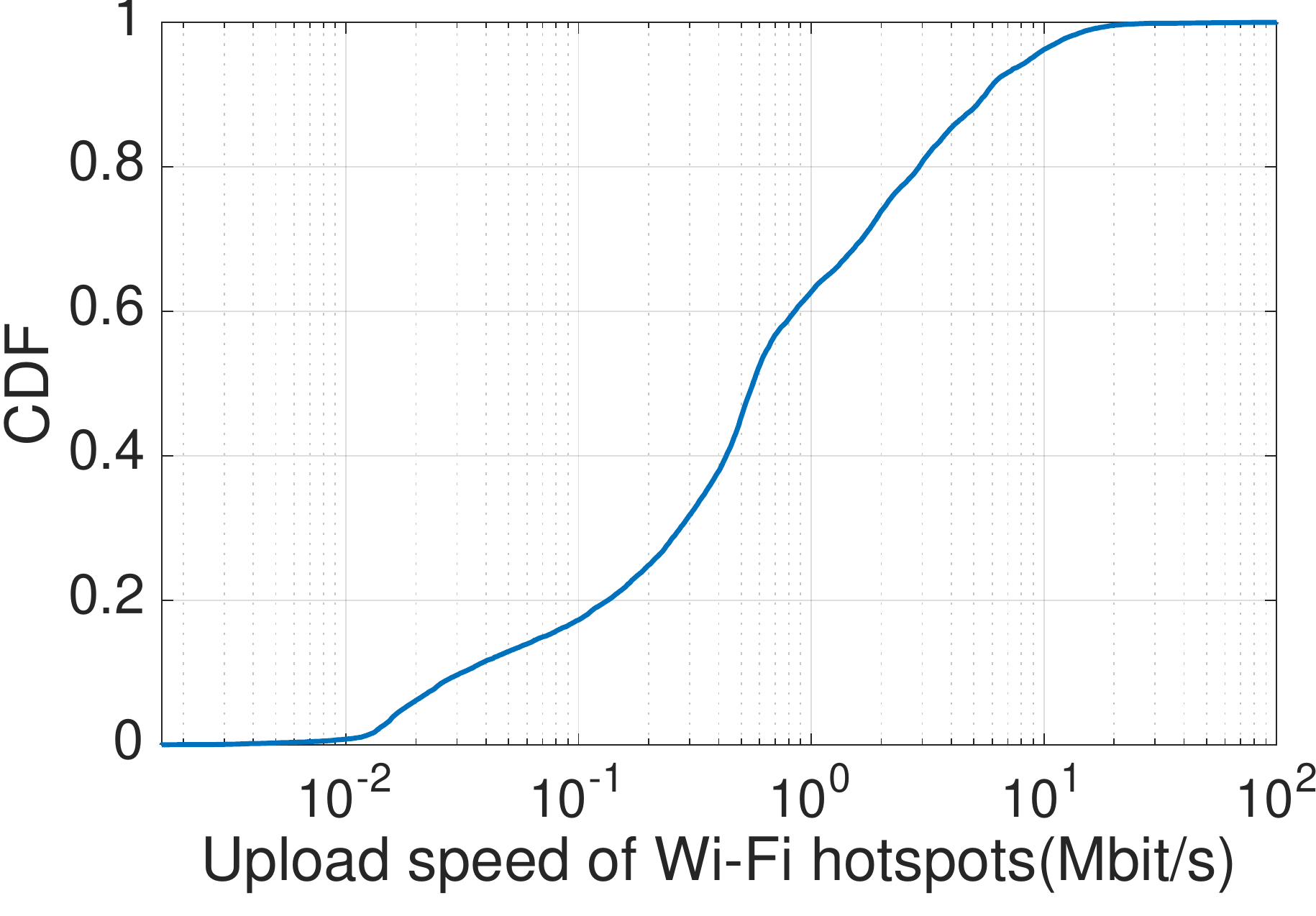}
        \caption{Upload bandwidth of Wi-Fi hotspots.}
        \label{fig:apBandwidthCdf}
    \end{minipage}
    \begin{minipage}{.333\textwidth}
        \centering
        \includegraphics[trim={0.5cm 7cm 2cm 8cm},clip,width=\linewidth]{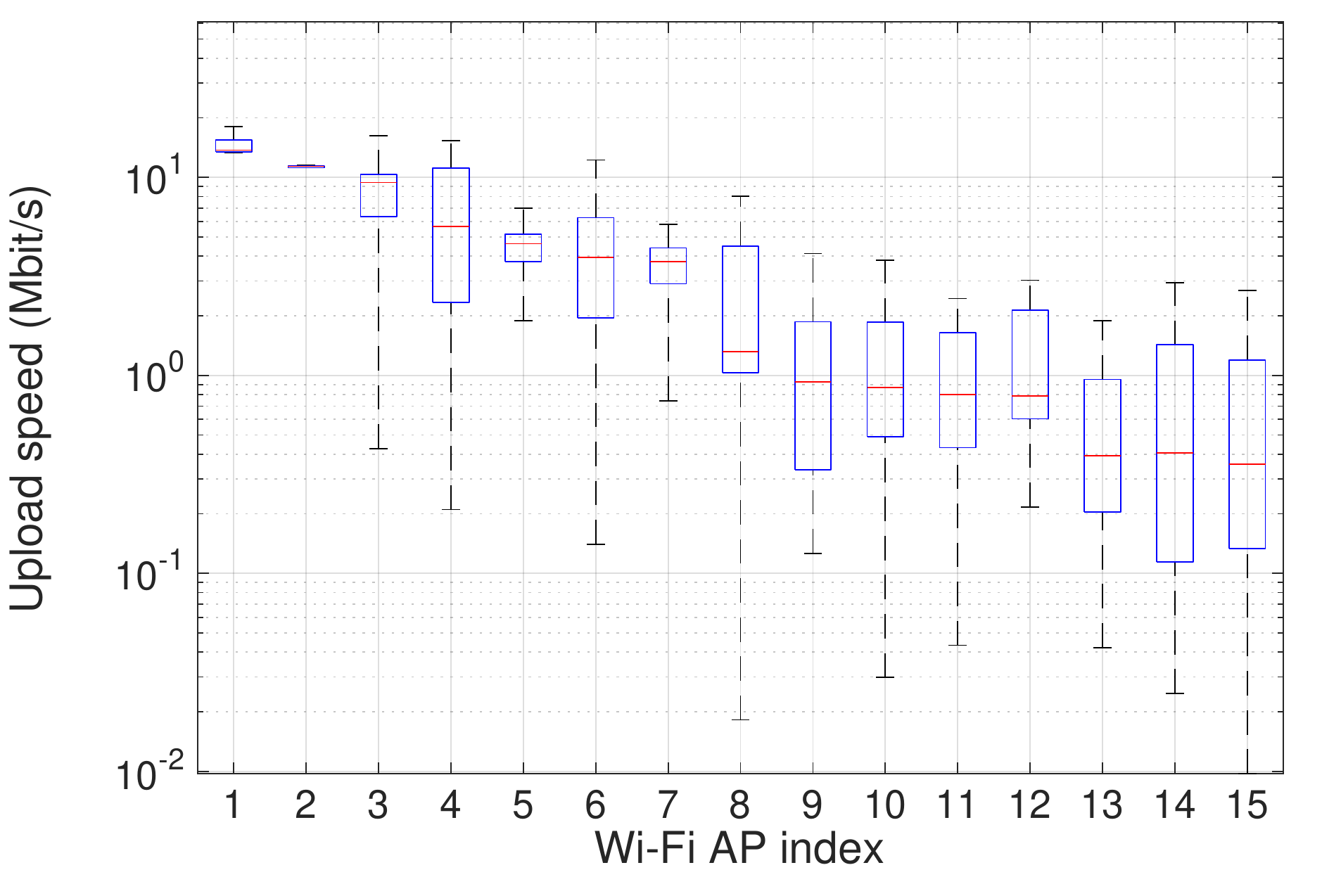}
        \caption{Upload speeds of Wi-Fi hotspots in the same cellular cell selected from a popular region.}
        \label{fig:apSpeed}
    \end{minipage}
\end{figure*}



By analyzing a massive data set collected by a crowdsourced Wi-Fi association mobile application which has more than $13$ million users, we observed that when users try to connect to a Wi-Fi hotspot, the chance of having more than $10$ other Wi-Fi networks around is over $50\%$. A promising idea to address this problem is to utilize all available edge networks simultaneously including Wi-Fi networks and cellular networks. Several previous efforts \cite{saeed2010dnis,habak2013optimal,habak2012g,tang2015application} have been devoted to using aggregated network resources to improve bandwidth capacity; however, they fail to increase the throughput for applications that originally use single-TCP connections to deliver data, e.g., most of the crowdsourced live streaming services.


In this paper, we propose \emph{BASS} (Bandwidth Aggregation SyStem), a multi-objective and deployable bandwidth aggregation system. The BASS system streams crowdsourced video data using \emph{aggregated} edge network resource via a portable device equipped with multiple network interfaces. BASS dynamically allocates \emph{aggregation servers} that receive video data from users and pass it to the original server, to enable MPTCP (i.e., Multi-path TCP \cite{ford2013tcp, barre2011experimenting, raiciu2012hard, honda2011still}) for today's representative crowdsourced live streaming platforms.

In our design of the BASS system, we try to fulfill the following requirements. (1) The system should be easy to deploy without any modifications to existing network infrastructure, servers and applications. (2) The design should be able to exploit multiple available edge networks including both Wi-Fi and cellular networks. (3) The strategies should dynamically allocate aggregation servers to provide satisfactory QoS for broadcasters.

%
%
%

Our contributions below provide a system design and strategies that satisfy these requirements.

First, we carry out measurement studies on $13$ million users connecting to $17$ million Wi-Fi hotspots in a time span of one month, to investigate (1) the availability of Wi-Fi resources in representative metropolitan areas (i.e., more than $50\%$ of the Wi-Fi sessions record that there are over $10$ other Wi-Fi hotspots around available), and (2) the insufficiency of most Wi-Fi hotspots' uplink bandwidths (i.e., over $60\%$ of the Wi-Fi sessions have an upload capacity lower than $1$ Mbit/s).

Second, we propose a practical aggregation network framework, which makes use of geo-distributed cloud resources for bandwidth aggregation for servers/users located at different places. Today's mobile devices usually do not support using multiple wireless interfaces simultaneously. As opposed to previous study \cite{habak2014oscar} that modifies the client devices, the adoption of middle-box and aggregation server in our design makes no modification to current operating system on mobile devices and makes \emph{BASS} more practical to be deployed for real-world devices and applications. Our design features: (1) The framework works transparently, i.e., broadcasters can use the original apps on their smart phones to upload high-bitrate video streams to the original servers. (2) Dynamical aggregation strategies, including i) a geo-distributed cloud aggregation server allocation algorithm that allocates cloud servers best matching the users and original servers (i.e., large bitrates can be achieved in the aggregation), and ii) a heuristic algorithm to maximize the whole system's overall bandwidth gain.

Third, we implement a prototype on EC2/PlanetLab to verify the effectiveness of our design. In particular, $8$ Amazon EC2 and $60$ Planet-Lab nodes are employed in our design to evaluate its performance. Compared with traditional TCP based uploading, our design can significantly improve the bandwidth by up to $5$ times.

The remainder of this paper is organized as follows. Section \ref{sec:measurement} presents our measurement results and motivation. Section \ref{sec:design} discusses the overall architecture of BASS system. Section \ref{sec:strategy} presents the dynamic allocation algorithm. In Section \ref{sec:evaluation}, we evaluate our algorithm via both simulations and experiments. Finally, Section \ref{sec:conclusion} concludes this paper and provides our plan for future work.

\section{Motivation: Measurement Studies} \label{sec:measurement}

\begin{figure*}
    \begin{minipage}{.3\textwidth}
        \vspace{4pt}
        \centering
        \includegraphics[width=0.9\linewidth]{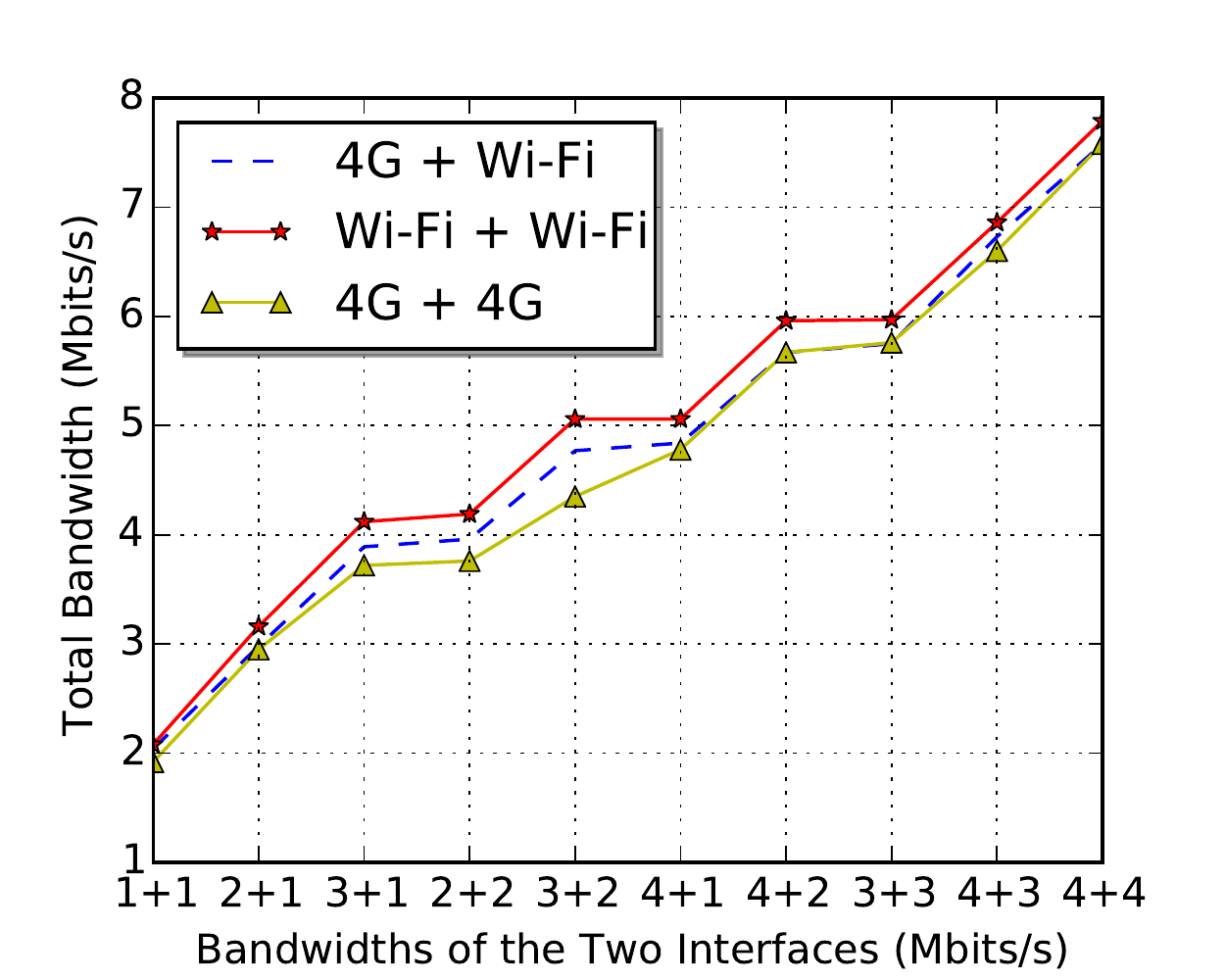}
        \vspace{4pt}
        \caption{Effectiveness of MPTCP for upload bandwidth aggregation.}
        \label{fig:bdTest}
    \end{minipage}
    \begin{minipage}{.7\textwidth}
        \centering
    	\includegraphics[width=\textwidth]{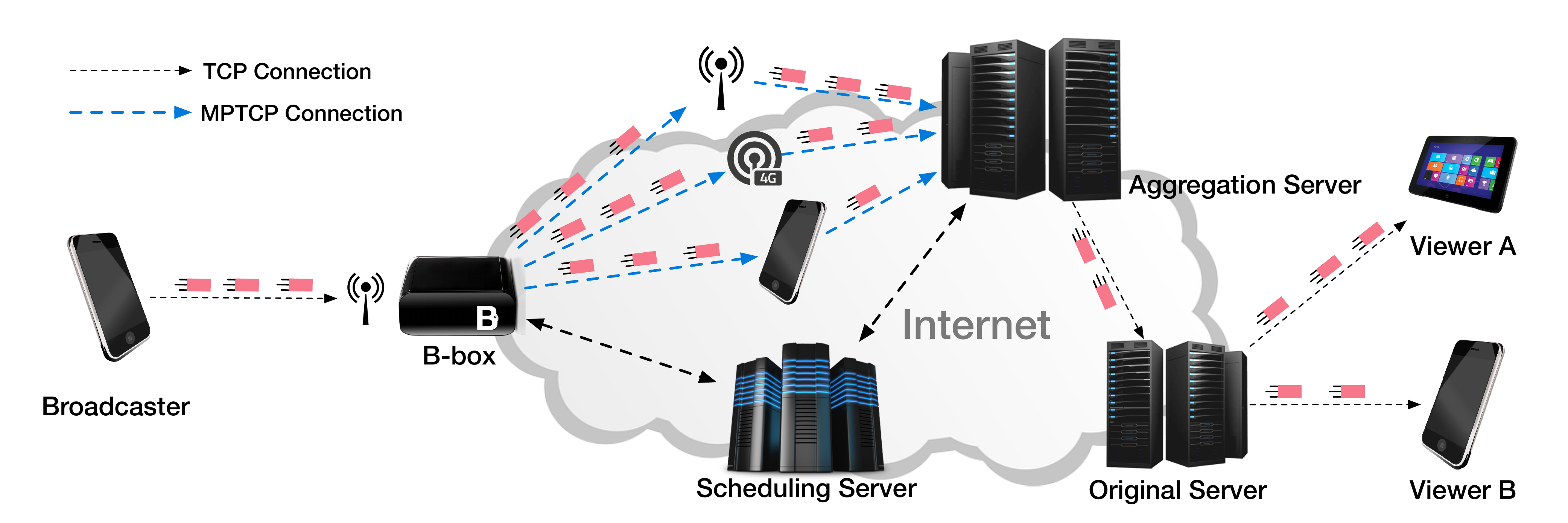}
        \caption{System architecture of BASS: Videos are uploaded using aggregated edge networks with portable B-box and cloud-based aggregation servers.}
        \label{fig:scenario}
    \end{minipage}
\end{figure*}
\subsection{Availability of Multiple Wi-Fi Networks}
To understand the availability of Wi-Fi resources, in this paper we analyzed a large dataset collected by a popular online crowdsourced Wi-Fi association mobile app in China. This app has over $200$ million downloads. It allows users to discover nearby crowdsourced free Wi-Fi hotspots. Every succeed connection is considered as a session, in which QoS statistics (e.g., measure of upload bandwidth of the Wi-Fi AP) will be reported to a trace server. In this paper, we use over $460$ million collected sessions of $13$ million users connecting to $17$ million Wi-Fi hotspots in a time span of one month to evaluate the availability of Wi-Fi resource.

In the traces collected, users report the available Wi-Fi hotspots around, based on which we are able to estimate the number of Wi-Fi hotspots available to a user. Fig. \ref{fig:aroundAp} presents the CDF of the number of Wi-Fi hotspots available to users. We observe that over $50\%$ of these sessions have more than $10$ Wi-Fi hotspots around users when they try to connect to a Wi-Fi network. This observation suggests that today's Wi-Fi deployment is highly promising for aggregation usage, i.e., using multiple Wi-Fi networks to aggregation the bandwidth capacity.

In Fig. \ref{fig:apBandwidthCdf}, we present the CDF plot of collected upload bandwidth test results. From this plot, we can observe that almost 60\% of collected bandwidth records are lower than 1 Mbit/s, which confirms that the bandwidth of today's edge network are very limited.

We next study the upload bandwidth of Wi-Fi hotspots inside a $100\times 100$m grid. Each sample shows the upload speed distribution of different APs located in the same cellular cell. From Fig. \ref{fig:apSpeed}, We observe significant difference of the median upload speeds among these APs, varying from 0.4 Mbit/s to 11 Mbit/s. The quality diversity in nearby hotspots suggests scheduling video chunk packets according to the status of the Wi-Fi interfaces is in demand.

\subsection{Bandwidth Improvement via MPTCP}




We verify the effectiveness of MPTCP in utilizing multiple wireless interfaces. MPTCP is a modification of the classical TCP that allows end-to-end data traffic to be split across multiple paths, while maintaining TCP connections at the end points (applications). Upper layers of the protocol stack only need to deal with a logical ``master'' TCP socket. Fig. \ref{fig:bdTest} illustrates a simple experiment conducted to measure the performance of MPTCP. We measured the bandwidth between local client and remote server under different network interface and bandwidth combinations. Both client and server are installed with MPTCP. The network interfaces we used in these experiments are two Wi-Fi interfaces and two 4G cellular interfaces. We observe that MPTCP is able to aggregate bandwidth resources from multiple networks.

To summarize, apart from the constant availability of cellular networks, users often have access to several Wi-Fi networks in urban areas, though these Wi-Fi networks often have limited bandwidth and heterogeneous upload capacities; also, MPTCP aggregation is promising to make full utilization of multiple wireless network resources for crowdsourced live stream uploading.

\section{System Components and Architecture of \emph{BASS}} \label{sec:design}

In this section, we provide a detailed description of the BASS system.
\subsection{System Components}

The BASS system has three main components. Firstly, a portable device B-box which can be treated as a special mobile Wi-Fi hotspot with several network interfaces and a built-in lightweight linux operating system (e.g., OpenWrt). Secondly, a group of aggregation servers with MPTCP enabled, usually allocated as cloud instances (e.g., Amazon EC2). Finally, a scheduling server which monitors aggregation servers' status and allocates aggregation servers to enable MPTCP.

\subsubsection{B-box: Bandwidth Aggregation Box}
B-box has a built-in lightweight linux operating system, equipped with several types of network interfaces including Wi-Fi and cellular. Thus, B-boxes are able to connect to multiple networks at the same time. The number of each kind of network interface is not fixed and can be adjusted according to user requirement when developed in the real world. B-box is deployed with MPTCP, therefore it can connect to a MPTCP-enabled server via multiple network interfaces. B-box also sets up a Wi-Fi hotspot using one of its Wi-Fi interfaces for users to connect. A B-box runs a proxy program that will receive video chunks uploaded by users via TCP and then upload them to the aggregation server via MPTCP. These proxy programs can be regular proxy programs like Shadowsocks, Squid or protocol converters\cite{detal2013multipath} designed specifically for MPTCP.


\subsubsection{Aggregation Server: Bandwidth Aggregation in the Cloud}

To avoid any modification to the crowdsource live streaming system and make BASS practical for today's crowdsource live streaming systems, we propose to allocate aggregation servers as a middleware to transfer data between original servers and B-boxes via MPTCP. Thus the data between B-box and the aggregation server can be delivered by several sub-flows via multiple paths. With bandwidth aggregation, B-box can obtain an aggregated bandwidth to provide a higher upload capacity for the users.

\subsubsection{Scheduling Server: Centralized Aggregation Scheduler}
Scheduling server is the coordinator between B-boxes and aggregation servers. Aggregation servers report their current loads (e.g., bandwidth consumed by served broadcasters) to scheduling server regularly. The scheduling server keeps track of the status and resource usages of all aggregation servers. When user starts a B-box, the B-box retrieves available aggregation server list from scheduling server and requests one aggregation server from scheduling server. Scheduling server assigns aggregation server to B-box and decides how much resource on this aggregation server will be allocated.


\subsection{Architecture}
Fig. \ref{fig:scenario} illustrates the architecture of BASS. In a nutshell, a broadcaster starts B-box which automatically requests aggregation server from scheduling server and sets up connection. Broadcaster uploads video streams to B-box. B-box then transmits the data via multiple interfaces to the aggregation server which will continually pass the data to the original server. The original servers are often dedicated streaming servers, for example, RTMP (Real Time Messaging Protocol over HTTP Tunnel) streaming servers deployed by Twitch.tv\cite{zhang2015crowdsourced}.

Since the original servers and users of a crowdsource live streaming system can be distributed in different locations, we propose to use a geo-distributed cloud solution to allocate aggregation servers at different locations accordingly, such that the aggregated bandwidth and latency can both be improved\cite{wuscaling2012}.
To optimize network performance and provide good latency and throughput performance, a deployment limited to 11 locations for North America or a total between 36 and 72 cloud-service locations with good peering connections on a global scale will be sufficient\cite{wang2011estimating}.


%

In our detailed design later, we will present how aggregation servers are allocated and how B-boxes are matched with different aggregation servers.




\section{Detailed Design of \emph{BASS}} \label{sec:strategy}

In this section, we present how aggregation servers are allocated and how they are matched with different B-boxes.

\subsection{Dynamical Aggregation Server Allocation}

In our design, the geo-distributed cloud solution allows us to deploy aggregation servers at many different locations around the world. When a B-box requests to send video data, an aggregation server is assigned to the B-box to aggregate video traffic. Aggregation servers are allocated dynamically according to both the user and original server's locations, the real-time bandwidth load of aggregation servers, and the network performance between the aggregation servers to the user and the original servers.
Considering these factors change over time, aggregation servers will be re-allocated regularly, e.g., every 30 minutes or when the network throughput on B-box is low.

We try to maximize the overall bandwidth for all users, while not exceeding the bandwidth capacities of aggregation servers. For example, when two B-boxes request the same aggregation server simultaneously, their original bandwidths of uploading to streaming server using single edge network and the increased bandwidths using aggregated edge networks are often different. Therefore the \emph{bandwidth gain} (i.e., the increased bandwidth comparing to the original bandwidth) are often different too.
Suppose this aggregation server's remaining bandwidth capacity can only serve one client, allocating this server to the client with the larger bandwidth improvement can yield more overall bandwidth gain.

Consider all these, we allocate aggregation servers to achieve the following objectives: (1) providing B-boxes with aggregation servers that can achieve high aggregated throughput; (2) matching aggregation servers with B-boxes to achieve maximum overall bandwidth gain.

\subsection{Algorithm and Implementation} \label{sec:algorithm}
We propose an allocation-matching algorithm to solve the problem above. We summarize important notations in Table \ref{tab:notation}. The allocation-matching algorithm is presented in Algorithm \ref{alg:algorithm}.

\begin{table}[!t]
\normalsize
\begin{center}
\caption{Important notations.} \label{tab:notation}
\begin{tabular}{p{.2\linewidth}|p{.7\linewidth}}
  \toprule
  Variable & Definition
  \\
  \midrule
  $n$ & Number of B-boxes requesting AS
  \\
  $m$ & Number of aggregation servers
  \\
  $t$ & Current time stamp
  \\
  $C$ & Set of clients
  \\
  $S$ & Set of aggregation servers
  \\
  $c_i$ & B-box $i$
  \\
  $s_j$ & Aggregation server $j$
  \\
  $B_o(c_i)$ & Bandwidth between $c_i$ and original server
  \\
  $B_o(s_j)$ & Bandwidth between $s_j$ and original server
  \\
  $B(c_i,s_j)$ & Bandwidth between $c_i$ and aggregation server $s_j$
  \\
  $B_o(c_i,s_j)$ & Bandwidth between $c_i$ and original server via $s_j$
  \\
  $R(s_j)$ & Remaining bandwidths capacity of $s_j$
  \\
  $T(s_j)$ & Total bandwidths capacity of $s_j$
  \\
  $r(s_j)$ & Bandwidth load rate of $s_j$
  \\
  $G(c_i, s_j)$ & Bandwidth gain B-box $c_i$ obtained from $s_j$
  \\
  $A(c_i, s_j)$ & 1 if $s_j$ is allocated to $c_i$, 0 otherwise
  \\
  $\Gamma_t$ & Overall bandwidth gain in time $t$
  \\
  \bottomrule
\end{tabular}
\end{center}
\end{table}

\begin{algorithm}[t!]
\caption{Aggregation and Matching Algorithm}\label{alg:algorithm}
\begin{algorithmic}[1]
\Procedure{Allocate Server}{}
\ForAll {$ c_i \in C$}
    \State \textit{find a subset of $S$, $S_o$, which is closest to $c$}
        \ForAll {$ s_j \in S_o$}
        \State \textit{Measure bandwidth $B(c_i,s_j)$, $B_o(c_i)$} 
        \State $B_o(c_i,s_j) \gets min\{ B(c_i, s_j), B_o(s_j)\}$
        \State $G(c_i, s_j) \gets B_o(c_i, s_j)-B_o(c_i)$
        \State \textit{Send bandwidth info to scheduling server}
    \EndFor
    \State \textit{Sort $G(c_i)$ in descending order}
\EndFor
\State \textit{Generate all possible allocation plans satisfy (\ref{equ:constraint})}
\State \textit{Calulate $\Gamma_t$ for each plan}
\State \Return \textit{allocation plan }$A$,
\Statex \hspace{\algorithmicindent}\textit{that achieves the highest bandwidth gain}
\State \textit{Update $R(s_j)$ for all aggregation server}
\EndProcedure
\end{algorithmic}
\end{algorithm}
Suppose a B-box $c_i$ requests for an aggregation server, it retrieves a list of available aggregation servers, uplink bandwidths from these aggregation servers to the original services. Bandwidths between the aggregation servers and the original servers are regularly updated by aggregation servers and report to the scheduling server. Scheduling server preferentially provides B-box with the list of aggregation servers that are close or have better connection to the original server.
We define bandwidth load rate $r(s_j)$ as follows:
$$
  r(s_j) = \frac{R(s_j)}{T(s_j)},
$$
Aggregation servers with $r(s_j)$ lower than a threshold will not be provided to a B-box. The threshold is determined by the bandwidth required to handle user's burst of throughput smoothly. Taking 4K streaming as an example, a stable 15 Mbits/s and max 50 Mbits/s bandwidth is enough to stream 4K. Therefore aggregation servers have to reserve some bandwidth in case of emergency. The threshold can be adjusted according to $T(s_j)$.

B-box then selects a subset $S_o$ of candidate aggregation servers which are in close proximity and measures $B(c_i,s_j)$, which is the uplink bandwidths to these aggregation servers via MPTCP. The bandwidth between $c_i$ and the original server $B_o(c_i)$ normally would be the largest bandwidth among each single edge network. Since without \emph{BASS}, users can only use one edge network.
The exact size of this subset and the time spent on measuring uplink bandwidth to each aggregation server is carefully adjusted to avoid adding too much overhead. For example, a B-box can select 3 aggregation servers and spend 5 seconds on measuring the bandwidth to each aggregation server to keep the overall start-up time relatively short.

These bandwidth test results including separate bandwidth on each path are stored in the B-box and updated regularly to avoid measuring bandwidth every time. After the B-box obtains the bandwidth information, it calculates $B_o(c_i,s_j)$, the predicted bandwidth to the original server via these aggregation servers, which is the smaller between the sum of many paths and the bandwidth from aggregation server to original server. Then the B-box reports the bandwidth test results to scheduling server.

Suppose the scheduling server receives $n$ requests from $n$ B-boxes in time $t$, then it will perform the matching between B-boxes and aggregation servers and try to find a matching that maximizes the overall bandwidth gain $\Gamma_t$, as follows:
\begin{align}
    & \text{maxmize } \Gamma_t = \sum_{i=1}^n\sum_{j=1}^m A(c_i, s_j)G(c_i, s_j), \\
    \text{subject to}& \notag \\
    & \sum_{i=1}^nA(c_i, s_j)B_o(c_i, s_j) < R(s_j),   j\in\{1,...,m\} \label{equ:constraint}
\end{align}
This problem can be solved in a centralized manner. Finally, the scheduling server will allocate aggregation servers according to the calculation and send results back to the B-boxes.



\section{Experiments and Performance Evaluation} \label{sec:evaluation}

\begin{figure*}
    \begin{minipage}{.33\textwidth}
        \centering
        \includegraphics[width=\linewidth]{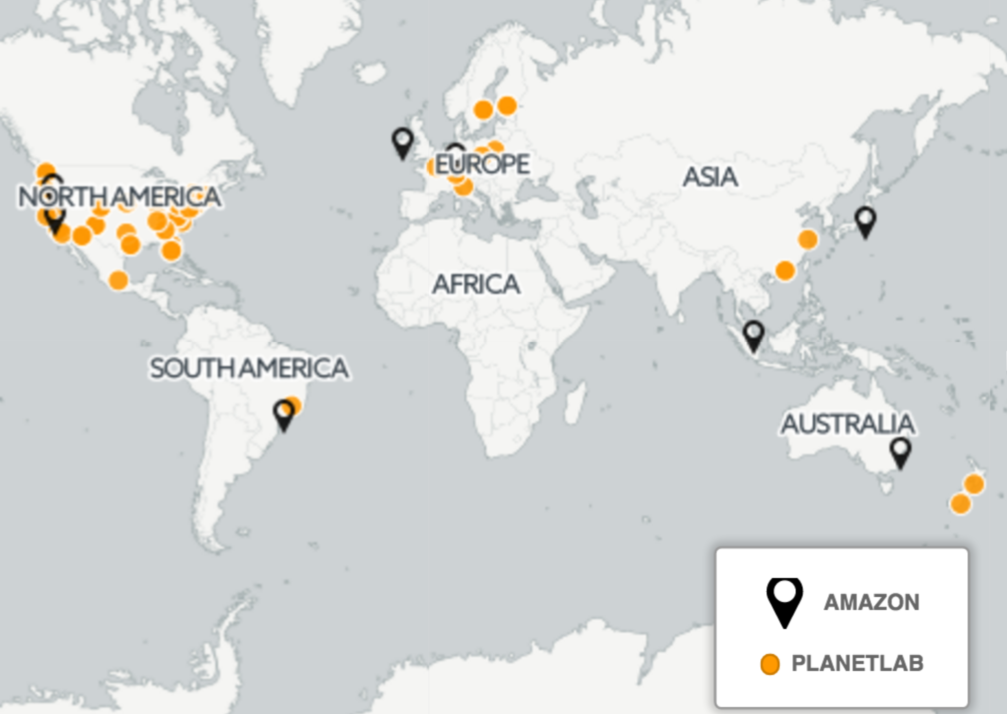}
        \caption{Locations of testing servers}
        \label{fig:serverdistribution}
    \end{minipage}
    \begin{minipage}{.33\textwidth}
        \centering
        \includegraphics[width=\linewidth]{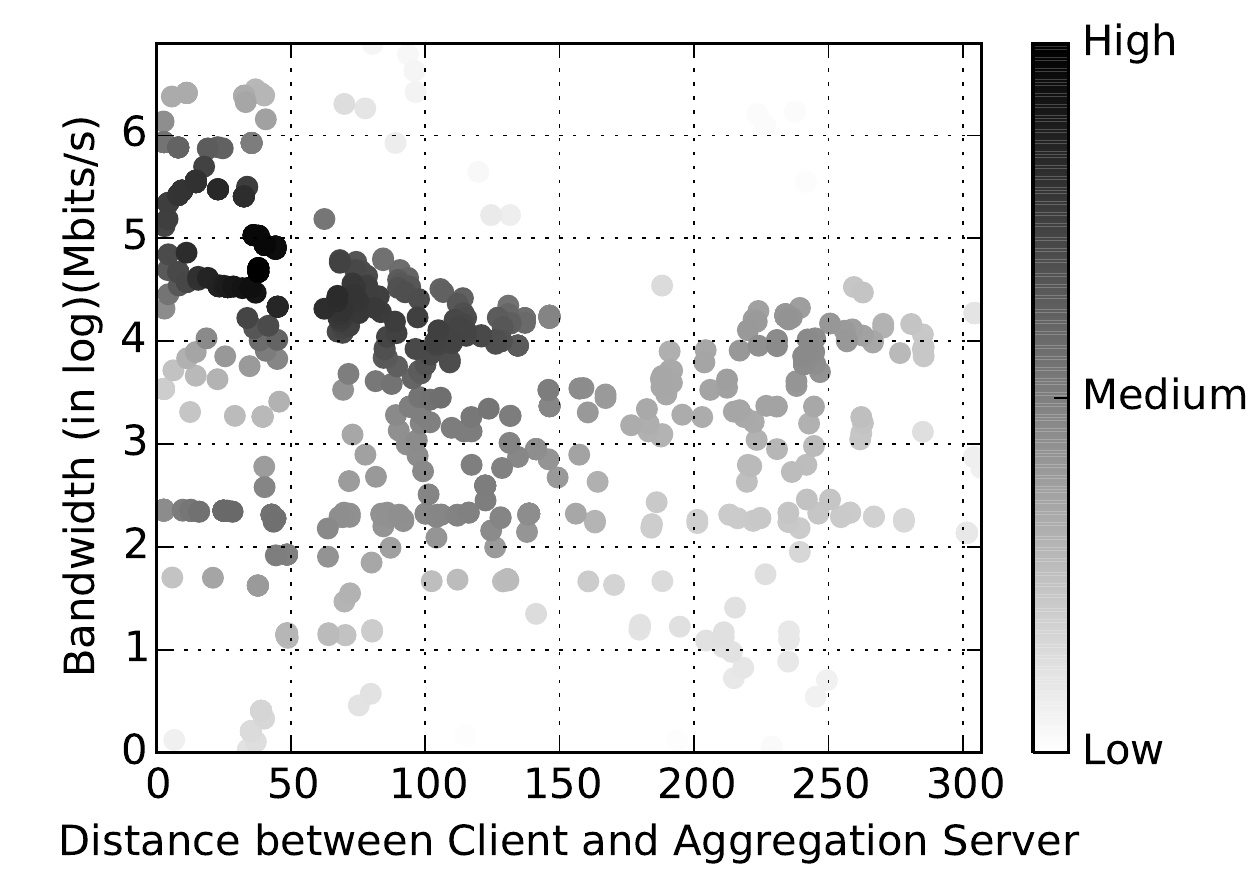}
        \caption{Relation between bandwidth and distance.}
        \label{fig:disandbw}
    \end{minipage}
    \begin{minipage}{.33\textwidth}
        \raggedright
        \includegraphics[width=\linewidth]{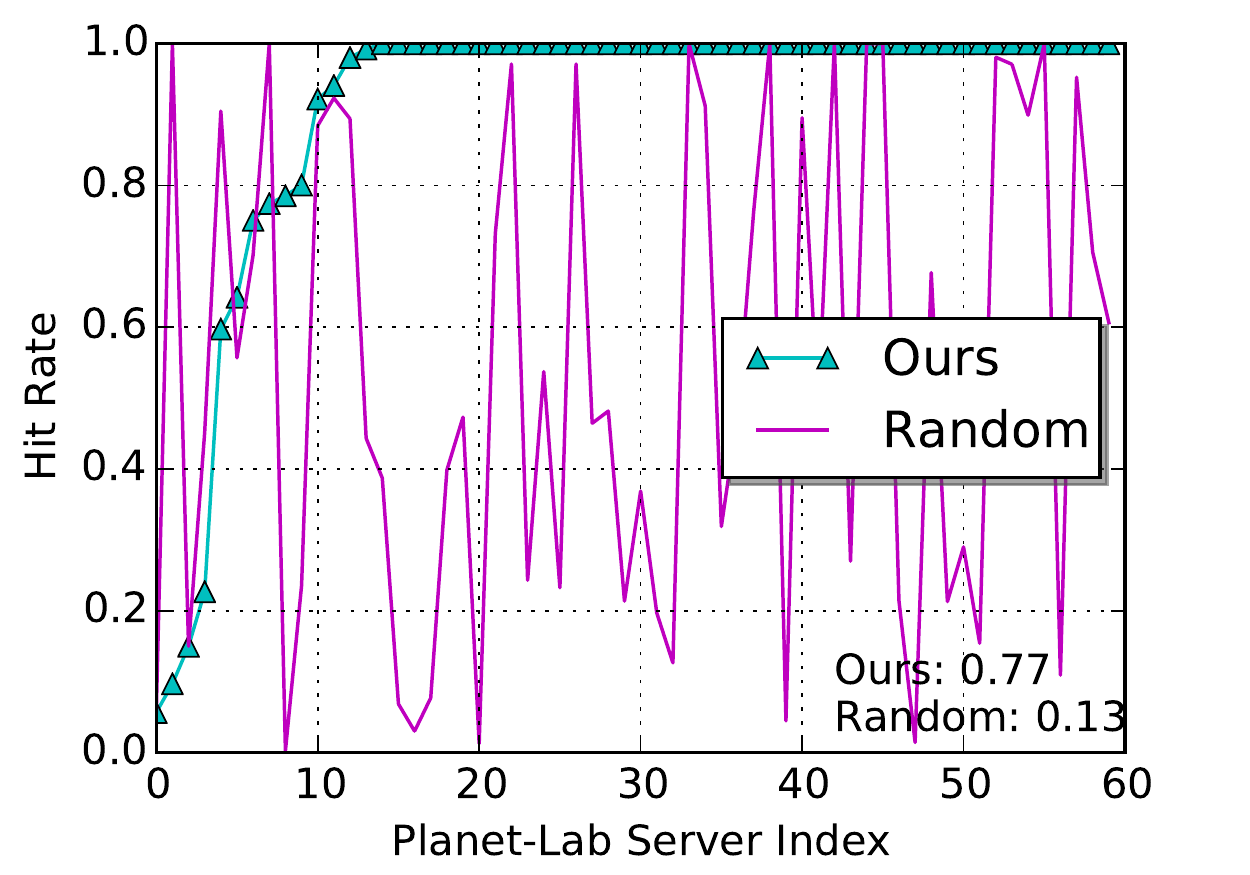}
        \caption{Comparison of aggregation server allocation.}
        \label{fig:allocationtest}
    \end{minipage}

    \vspace{20pt}
    \begin{minipage}{.33\textwidth}
        \centering
        \includegraphics[width=\linewidth]{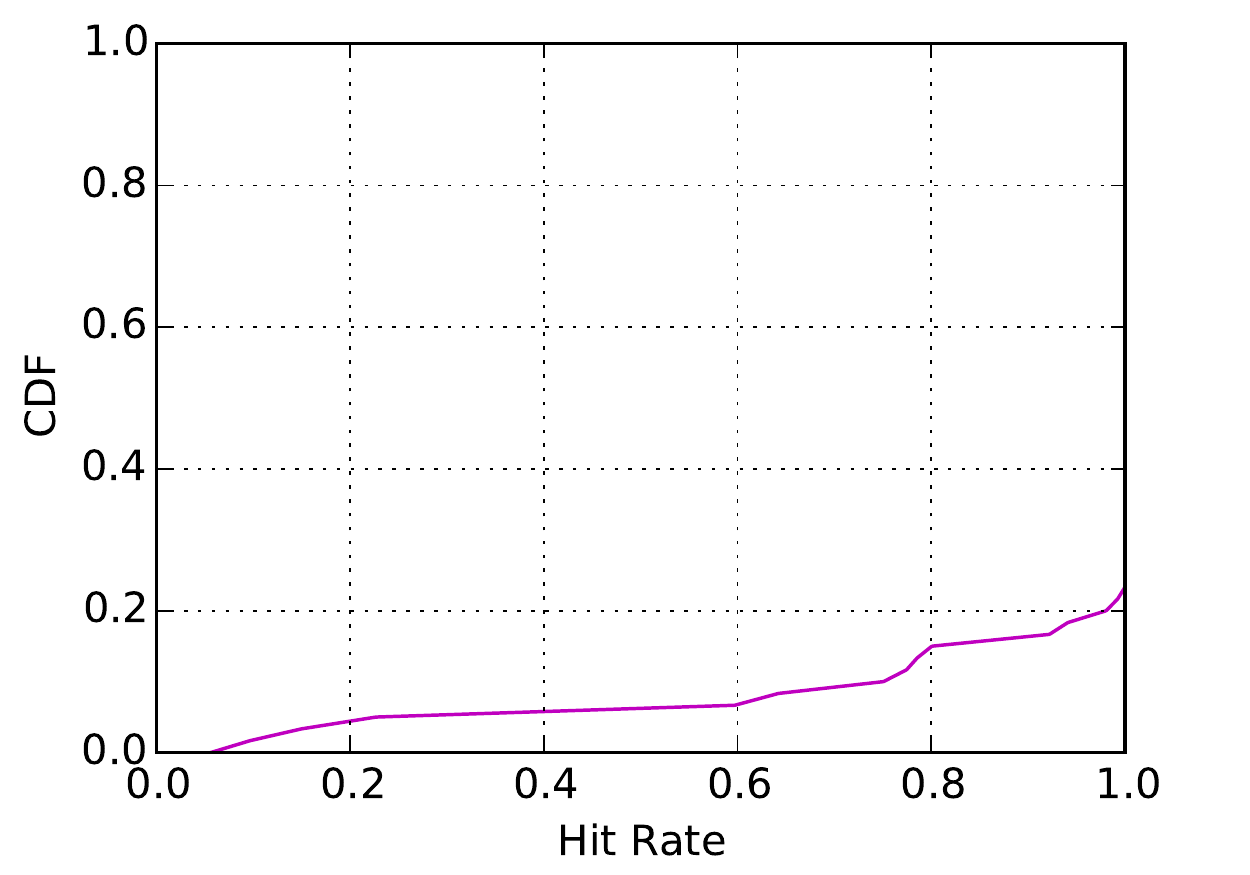}
        \caption{CDF of Hit Rate $\gamma$}
        \label{fig:cdf}
    \end{minipage}
    \begin{minipage}{.33\textwidth}
        \centering
        \includegraphics[width=\linewidth]{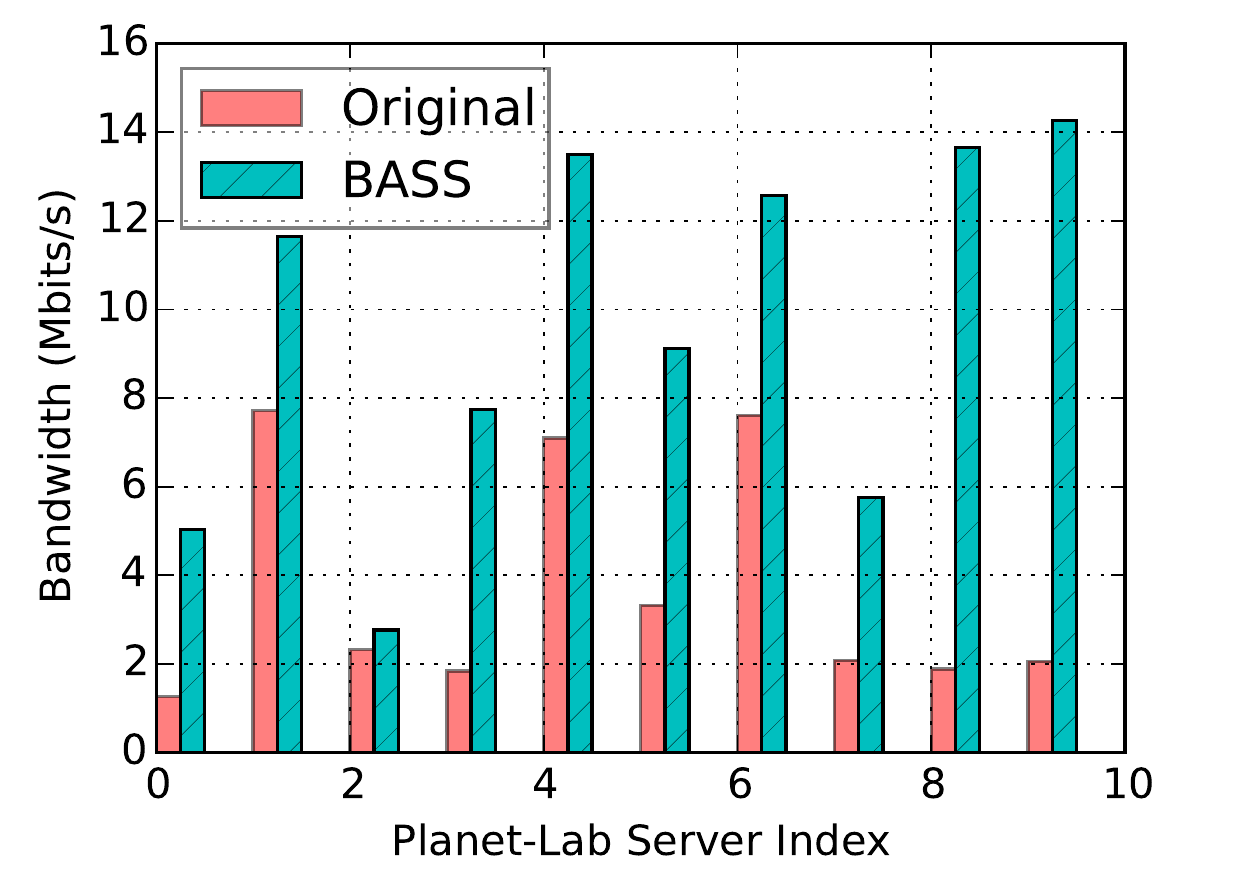}
        \caption{Bandwidth achieved at a local client }
        \label{fig:realworldTest}
    \end{minipage}
    \begin{minipage}{.33\textwidth}
        \centering
        \includegraphics[width=\linewidth]{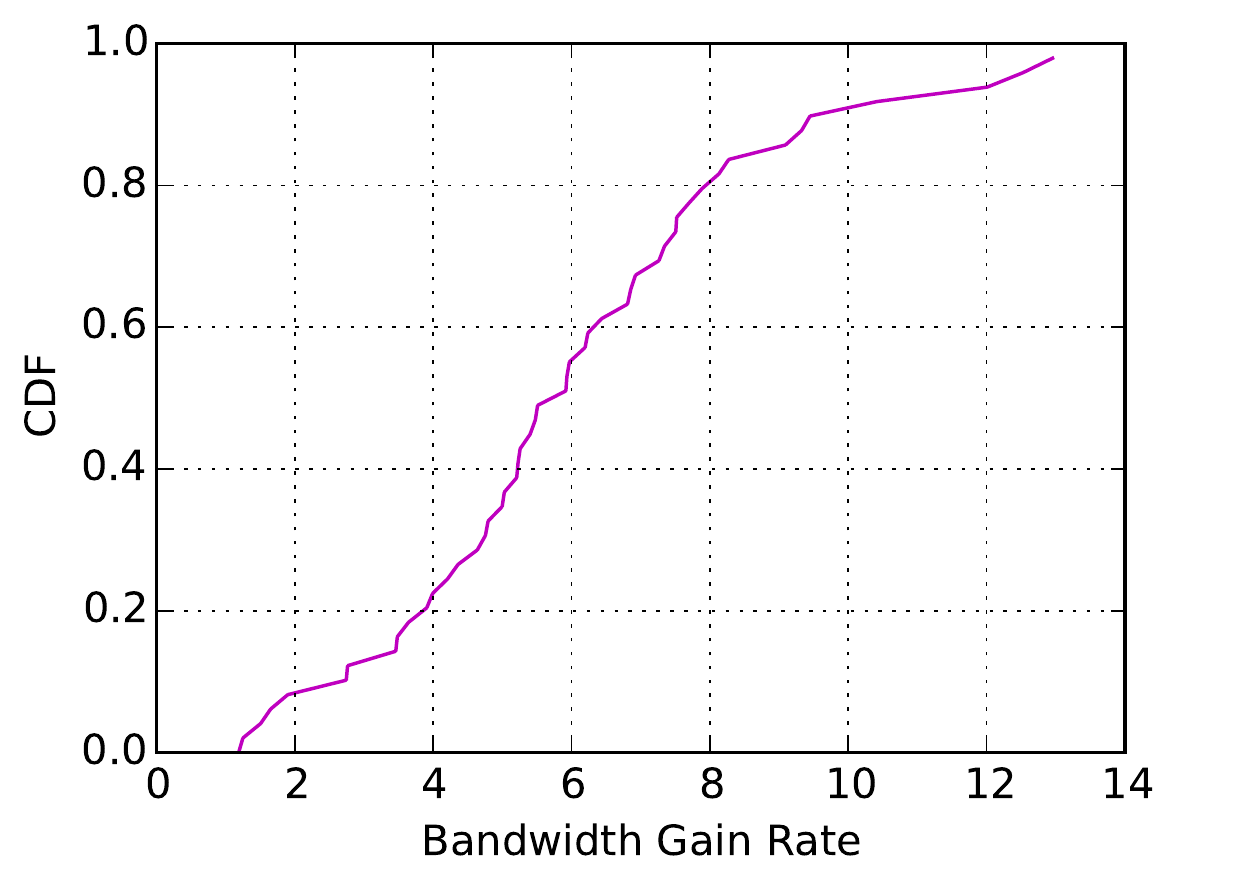}
        \caption{CDF of bandwidth gain.}
        \label{fig:bdgaincdf}
    \end{minipage}
\end{figure*}
In this section, we evaluate the performance of our system strategies by real-world experiments on $8$ Amazon EC2 instances and $60$ Planet-Lab nodes. We begin with the evaluation of the feasibility of selecting nearby aggregation servers to provide high network throughput. Then we evaluate the dynamical allocation of aggregation servers.

\subsection{Experiment Setup}

In our experiments, we implemented a B-box prototype on a Raspberry Pi 2 device with kernel version 3.18, which has two Wi-Fi interfaces and runs our own proxy program.
We use Amazon EC2 nodes (micro type) to install MPTCP and act as candidate aggregation servers. We use $60$ Planet-Lab nodes to act as B-boxes or original servers. The locations of these servers are shown in Fig. \ref{fig:serverdistribution}. Each B-box will be scheduled to upload video data to a randomly selected original server, and an aggregation server will be selected to perform the aggregation and traffic proxy. We deploy bandwidth measurement tools including \textsf{iperf} on the EC2 and Planet-Lab nodes and run the measurements using a Python script.

\subsection{Experiment Results}

\subsubsection{Effectiveness of Distance-based Candidate Aggregation Servers}

In Fig. \ref{fig:disandbw}, we plot the bandwidth between B-boxes and aggregation servers versus the distance between them. We observe that most PlanetLab servers that obtained higher bandwidth are closer to the Amazon EC2 servers, indicating that the candidate aggregation servers selection step is promising to choose proper aggregation servers.

\subsubsection{Effectiveness of the Aggregation Server Selection}

We compare our algorithm with a random aggregation server selection algorithm, in which aggregation servers are randomly assigned to B-boxes. We define the hit rate $\gamma$ when selecting an aggregation server as follows:
$$
  \gamma = \frac{b}{B},
$$
where $b$ is the bandwidth achieved using this aggregation server, and $B$ is the optimal bandwidth that can be achieved among all the aggregation servers. $\gamma=1$ if the algorithm selects the aggregation server with highest bandwidth. The experiment result is shown in Fig. \ref{fig:allocationtest}. We observe that our algorithm can find the best aggregation servers for the B-boxes and provide the maximum bandwidth most of time. We further plot the CDF of the hit rates in our design in Fig. \ref{fig:cdf}. We observe that for over $77\%$ of the tests, the hit rate $\gamma$ is 1, indicating that our server allocation performs as well as the optimal solution.

\subsection{Bandwidth Experienced by Clients}

Finally, we set up a local PC as the crowdsourced streaming client, which uploads data via B-box prototype. The bandwidths of both interfaces are limited to $8$Mits/s. We first measure the bandwidth between client to all possible original servers (i.e., the $60$ Planet-Lab nodes). Then we measured the bandwidths achieved by using our BASS system. As illustrated in Fig. \ref{fig:realworldTest}, the bars are bandwidth achieved in the original TCP paradigm and in BASS, respectively. We observe that BASS system can significantly improve the network performance. The bandwidth can be improved by up to $14$ times.

In particular, we measure the improvement in all the tests. We plot the CDF of bandwidth improvement in Fig. \ref{fig:bdgaincdf}. The bandwidth gain rate varies between $1.14$ and $14.59$ with an average of $5.11$. For over $50\%$ of the tests, an improvement of $6$X is achieved in our design.

\section{Conclusion} \label{sec:conclusion}

With the launch of new crowdsourced live streaming service like Facebook live, its popularity keeps soaring. In this paper, we proposed BASS, a deployable edge network bandwidth aggregation system for improving the uplink bandwidths of broadcasters.
We first presented our measurement studies based on a extensive dataset collected by a popular Wi-Fi association mobile app in China. We then presented the BASS architecture, designed an aggregation server allocation strategies, and implemented a prototype on a Raspberry Pi 2 device and Amazon EC2 servers. We evaluated the feasibility and performance on Planet-Lab testbed. The results show stable and promising performance gain.
For future work, we plan to implement our system for real-world services, and evaluate its performance in the wild. We plan to (1) extend our system by adding more functions including scheduling video chunk packets according to path characteristics, (2) optimize our system for multiple objectives including cost, Wi-Fi interference and energy consumption.

\bibliographystyle{IEEEbib}
\bibliography{main}


\end{document}